\documentclass[aps,graphicx,twocolumn]{revtex4}

\usepackage{graphicx}
\usepackage{amsmath}

\begin{document}

\title{Entanglement purification of nonlocal quantum-dot-confined electrons assisted by double-sided optical microcavities\footnote{Published in  Ann. Phys. (Berlin)  \textbf{530}, 1800029 (2018).}}

\author{Zi-Chao Liu$^{1,2}$, Jian-Song Hong$^{1,2}$, Jia-Jie Guo$^{1,2}$,
Tao Li$^{1,3,}$\footnote{Corresponding author: tao.li.fv@riken.jp},
Qing Ai$^{1}$, Ahmed Alsaedi$^{2}$, Tasawar Hayat$^{2,4}$, and
Fu-Guo Deng$^{1,2,}$\footnote{Corresponding author:
fgdeng@bnu.edu.cn}}

\affiliation{$^{1}$Department of Physics, Applied Optics Beijing
Area Major Laboratory,
Beijing Normal University, Beijing 100875, China\\
$^{2}$NAAM-Research Group, Department of Mathematics, Faculty of
Science, King Abdulaziz University,  Jeddah 21589, Saudi Arabia\\
$^{3}$CEMS, RIKEN, Wako-shi, Saitama 351-0198, Japan\\
$^{4}$Department of Mathematics, Quaid-I-Azam University, Islamabad
44000, Pakistan}

\date{\today }

\begin{abstract}
We present a nondestructive parity-check detector (PCD) scheme for
two single-electron quantum dots  embedded in double-sided optical
microcavities. Using a polarization-entangled photon pair, the PCD
works in a parallel style and is robust to the phase fluctuation of
the optical path length. In addition, we present an economic
entanglement purification protocol for electron pairs with our
nondestructive PCD. The parties in quantum communication can
increase the purification efficiency and simultaneously decrease the
quantum source consumed for some particular fidelity thresholds.
Therefore, our protocol has good applications in the future quantum
communication and distributed quantum networks.
 \end{abstract}

\maketitle

\section{Introduction}


Quantum communication is a very promising technique for sending
information in an absolutely secure way \cite{Quaninternet}. Quantum
entanglement, as one of the core resources of quantum information,
plays an important role in quantum communication, quantum metrology,
and distributed quantum computation
\cite{nielsen2000,QMee,DisQC1,DisQC2,DisQC3,DisQC33,DisQC4}.
However, quantum entanglement could only be created locally and the
distribution of quantum entanglement is  a prerequisite for various
quantum communication protocols, such as quantum key distribution
\cite{ekert1991,bennett1992a,LiaoPanQKD,Lo2014QKD}, quantum
teleportation \cite{telep1,RenPanteleportation}, quantum secret
sharing \cite{QSS0,QSS1,QSS2}, and  quantum secure direct
communication \cite{QSDC1,QSDC2,QSDC3,QSDC6,QSDC5,QSDC7}. The
distance of entanglement distribution, in practice, is largely
limited by channel noise, leading to an exponential attenuation and
decoherence when increasing the channel distance \cite{nielsen2000}.

Quantum repeater \cite{QR0,QR1,QR2} is an efficient protocol to
establish a long-distance entanglement channel connecting two remote
quantum nodes in a quantum communication network. By dividing the
channel into many short-distance channels with several quantum
nodes, the locally created entanglement will be shared by
neighboring nodes after entanglement distribution. The entanglement
is then further extended to cover larger distance by entanglement
swapping \cite{QS0,QS1,QS2,QS3}. The decrease of the entanglement
due to  channel noise could be compensated by  quantum entanglement
concentration
\cite{ECP1B,ECP1add1,ECP1add2,Eccs1,Eccs2,ECP1add3,Eccl1,Eccl2} and
quantum entanglement purification
\cite{EPP1B,Qerrer,EPPpan01,EPPpan2,EPPSheng1,EPPpdeng1}, which in
principle could lead to the maximal entanglement between any two
nodes in a quantum communication network \cite{Quaninternet}.

Entanglement purification is one of the passive methods to depress
the negative effects of channel noise \cite{Qerrer}. It can obtain
some high-fidelity nonlocal entangled subsystems from an ensemble in
a mixed entangled state. In 1996, Bennett \emph{et al} \cite{EPP1B}
presented the first entanglement purification protocol (EPP) to
purify a Werner state with quantum controlled-not (CNOT) gates and
bilateral rotations. In 2001, Pan \emph{et al} \cite{EPPpan01}
introduced a simplified EPP with polarization beam splitters and
single-photon detectors. In 2002, Simon and Pan \cite{EPPpan2}
proposed an EPP for a practical parametric down-conversion source.
Especially, based on the cross-Kerr nonlinearities, an efficient
polarization EPP and an efficient multipartite EPP were proposed by
Sheng \emph{et al} \cite{EPPSheng1}  and Deng \cite{EPPpdeng1},
respectively. In these schemes, one can increase the fidelity of
quantum states by repeatedly performing the purification protocols
\cite{EPPSheng1,EPPpdeng1}. In 2010, Sheng and Deng \cite{DEPP1}
proposed the concept of deterministic entanglement purification for
two-photon entangled systems, and they presented a two-step
deterministic EPP for polarization entanglement with the
hyperentanglement. Subsequently, Li \cite{DEPP2} and Sheng and Deng
\cite{DEPP3} independently proposed a one-step deterministic EPP for
polarization entanglement with only the spatial entanglement of
photon pairs, resorting to linear-optical elements. These
deterministic EPPs are extended to purify multipartite entanglement
\cite{DEPP4} and to the one assisted by time-bin entanglement
\cite{DEPP5}. Subsequently, Sheng and Zhou \cite{EPPsl2} proposed a
deterministic EPP for secure double-server blind quantum
computation. Deterministic EPPs
\cite{DEPP1,DEPP2,DEPP3,DEPP4,DEPP5,EPPsl2} are significantly
different from conventional EPPs
\cite{EPP1B,Qerrer,EPPpan01,EPPpan2,EPPSheng1,EPPpdeng1} as they
work in a completely deterministic way, which  largely reduces
quantum resources in a quantum network. The another interesting
branch in the development of EPP is the hyperentanglement
purification. In 2013, Ren and Deng \cite{HEPP1} presented the first
hyperentanglement purification protocol (hyper-EPP) for two-photon
systems in polarization-spatial hyperentangled states, and it is
very useful in  high-capacity quantum networks. In 2014, Ren, Du,
and Deng \cite{HEPP2} gave a two-step hyper-EPP for
polarization-spatial hyperentangled states with the
quantum-state-joining method, and it has a far higher efficiency. In
2016, Wang, Liu, and Deng \cite{HEPP3} presented a general hyper-EPP
for two-photon six-qubit quantum systems, and it has a even higher
capacity. Besides, there are also some other interesting EPPs
\cite{EPPHybrid,EPPKerr,DEPPw1,DEPPw2,DEPPw3,DEPPw4}, such as the
first hybrid EPP for discrete-variable and continuous-variable
entanglement \cite{EPPHybrid}, the EPP using cross-Kerr nonlinearity
by identifying the intensity of probe coherent beams \cite{EPPKerr},
and so on.

%


Singly-charged quantum dots (QD) are one of the competitive
candidates for quantum information processing
\cite{QD00,QD001,QD02}, due to their unique character for
implementing qubits and the mature developed semiconductor
technology \cite{QD01}.  Fast manipulation and measurement on single
QD have also been experimentally demonstrated \cite{QD03}.
Meanwhile, the electron-spin coherence time of the QD  is
demonstrated to be the scale of several microseconds \cite{QD04},
and it is long enough when compared with the picoseconds
manipulating time on single QDs. By embedding a singly-charged
negative QD in a microcavity, one can implement an interesting
interface and achieve the effective interaction between an electron
spin and a single photon. In 2008, Hu \cite{QD1} detailed the giant
circular birefringence originated from the spin-dependent dipole
coupling for a negatively charged QD embedded in a micropillar
cavity. This QD-cavity system is then extensively researched and is
exploited for photon-photon and spin-photon entanglement generation
\cite{QD2,QD.2,QD22}, universal  quantum gates
\cite{QD3,QD4,QD42,QD41,QDreview}, Bell-state analyzers
\cite{QD50,QD5,QD5add1,QD5add2}, and quantum transistors and routers
\cite{QDrout1,QDrout2}. Furthermore, a general EPP for electron-spin
entangled states was proposed by constructing a parity-check
detector (PCD) for two electron spins \cite{QDpcd1}. In this EPP,
the PCD is implemented by subsequently scattering one single photon
with two QD-cavity unions and two effective input-output processes
are involved. Recently, Li and Deng \cite{QDpcd2} proposed an
error-rejected PCD for electron spins with one effective
input-output process, and it is robust to various kinds of errors
involving in a practical single-photon scattering process.

In this paper, we propose a high-performance PCD for QD-confined electron-spin systems based on linear optical elements and double-sided microcavities. By introducing a polarization-entangled photon source and inputting each photon of the entangled photon pair into the corresponding QD-cavity unions, the PCD works in a parallel way and is heralded by the coincident click of single-photon detectors. The two photons are scattered simultaneous by the corresponding QD-cavity union, and then they are measured immediately after the scattering process other than that measured in the middle point after a single-photon interference \cite{QDpcd2}. Therefore, our parity-check detector will be more efficient than the ones based on single-photon scatting process when dealing with nonlocal electron spins that depart from each other. In addition, we present an economic EPP for electron-spin pairs with our PCD. We can increase the purification efficiency and simultaneously decrease the quantum source consumed for some particular fidelity thresholds, due to the purification-order independent character of the EPP.

\section{PARITY-CHECK Detector based on DOUBLE-SIDEd MicroCAVITY}

\begin{figure}[!tpb]
  \centering
  \includegraphics[width=8.5cm]{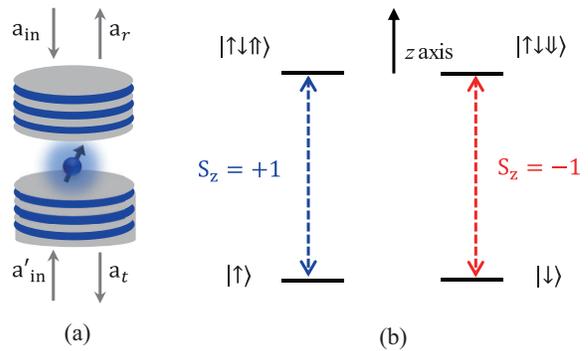}\\
  \caption{(a) Schematic diagram for a QD embedded in the double-sided microcavity. (b) Energy-level diagram for a negatively charged single-electron QD. }\label{fig1}
\end{figure}

The QD-cavity union consists of a double-sided optical cavity and a negatively charged single-electron QD is shown in Fig.\ref{fig1} (a) \cite{QD2,QD.2}. When the QD is excited, it creates a negatively charged trion $X^-$ that consists of two electrons and a positive hole. There are two degenerate states of the $X^-$ denoting as $|\!\!\uparrow\downarrow\Uparrow\rangle$ and $|\!\!\downarrow\uparrow\Downarrow\rangle$. According to Pauli's exclusion principle,  two possible transitions of the QD are respectively related to the creation or annihilation of a single polarized photon of spin $s_{z}=+1$  and $s_{z}=-1$, shown in Fig.\ref{fig1} (b). Ideally, for a particular input state, the output state after the scattering process will be detailed as follows \cite{QD2,QD.2}:
\begin{equation}\label{mcf}
  \begin{aligned}
&|R^{\uparrow},\uparrow\rangle\rightarrow|L^{\downarrow},\uparrow\rangle,&
&|L^{\uparrow},\uparrow\rangle\rightarrow-|L^{\uparrow},\uparrow\rangle,&\\
&|R^{\downarrow},\uparrow\rangle\rightarrow-|R^{\downarrow},\uparrow\rangle,&
&|L^{\downarrow},\uparrow\rangle\rightarrow|R^{\uparrow},\uparrow\rangle,&\\
&|R^{\uparrow},\downarrow\rangle\rightarrow-|R^{\uparrow},\downarrow\rangle,&
&|L^{\uparrow},\downarrow\rangle\rightarrow|R^{\downarrow},\downarrow\rangle,&\\
&|R^{\downarrow},\downarrow\rangle\rightarrow|L^{\uparrow},\downarrow\rangle,&
&|L^{\downarrow},\downarrow\rangle\rightarrow-|L^{\downarrow},\downarrow\rangle.&
  \end{aligned}
\end{equation}
Here $|p,s\rangle=|p\rangle\otimes|s\rangle$ denotes a product state of a single photon $|p\rangle$ and an electron spin $|s\rangle$, for example $|R^{\uparrow},\uparrow\rangle=|R^{\uparrow}\rangle\otimes|\uparrow\rangle$. The states $|\uparrow\rangle$ and $|\downarrow\rangle$ represent spin-up and spin-down states of the electron. $|L\rangle$ and $|R\rangle$ represent right-circularly polarized and left-circularly polarized single photons along the propagation direction, respectively. The superscript arrow $\uparrow$ or ${\downarrow}$ represents that the direction of the input photon is parallel or antiparallel to the $z$ axis of the electron spin.

\begin{figure}[!tpb]
  \centering
  \includegraphics[width=8.5cm]{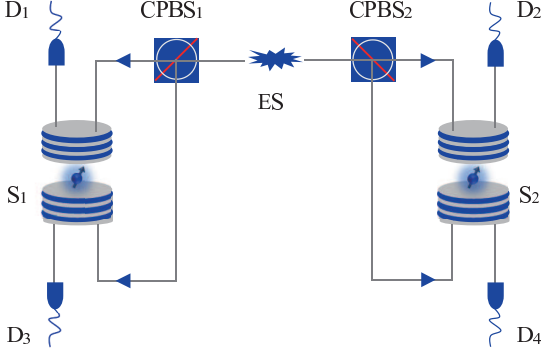}\\
  \caption{Schematic diagram for the nondestructive PCD. CPBS$_i$, ($i=1, 2$)  is the circularly-polarized beam splitter that transmits photons in state $|R\rangle$ and reflects photons in state $|L\rangle$. $D_{j}$ $(j=1, 2,...)$ represents four single-photon detectors. ES represents an entanglement source that emits  photon pairs in polarization-entangled state $|\Phi_+\rangle=(|RR\rangle+|LL\rangle)/\sqrt{2}$. }\label{qnd}
\end{figure}

Based on the conditional scattering detailed in Eq. (\ref{mcf}), we propose a nondestructive PCD scheme for measuring the parity of  two-electron systems. The principle of our PCD is shown in Fig. \ref{qnd}. The entanglement source (ES) emits photon pairs in the maximal entangled state $|\Phi_+\rangle=(|RR\rangle+|LL\rangle)/\sqrt{2}$. The two photons are, respectively, injected into two identical QD-cavity unions. In each side, the photon is firstly divided into two propagating modes of different polarizations by a circularly polarized beam splitter (CPBS), and then the two modes are transmitted into the microcavity from each side. After the interaction between the photon and electron, the two photons will be output from either the top or bottom of the corresponding microcavity and click two of the four single-photon detectors. When the spin parity of the electrons is even, detectors of the same side $D_1D_2$ or $D_3D_4$ will click. However, when the parity is odd, detectors of the opposite sides $D_1D_3$ or $D_2D_4$ will click. To be specific, the evolution of this process can be detailed as follows:
\begin{equation}\label{gate1}
  \begin{split}
  \frac{1}{\sqrt{2}}(|R^{\downarrow}R^{\downarrow}\rangle+|L^{\uparrow}L^{\uparrow}\rangle)|\!\!\uparrow\uparrow\rangle
  \rightarrow\frac{1}{\sqrt{2}}(|R^{\downarrow}R^{\downarrow}\rangle+|L^{\uparrow}L^{\uparrow}\rangle)|\!\!\uparrow\uparrow\rangle,\\
  \frac{1}{\sqrt{2}}(|R^{\downarrow}R^{\downarrow}\rangle+|L^{\uparrow}L^{\uparrow}\rangle)|\!\!\downarrow\downarrow\rangle
  \rightarrow\frac{1}{\sqrt{2}}(|L^{\uparrow}L^{\uparrow}\rangle+|R^{\downarrow}R^{\downarrow}\rangle)|\!\!\downarrow\downarrow\rangle,
  \end{split}
\end{equation}
and
\begin{equation}\label{gate2}
  \begin{split}
  \frac{1}{\sqrt{2}}(|R^{\downarrow}R^{\downarrow}\rangle+|L^{\uparrow}L^{\uparrow}\rangle)|\!\!\uparrow\downarrow\rangle
  \rightarrow-\frac{1}{\sqrt{2}}(|R^{\downarrow}L^{\uparrow}\rangle+|L^{\uparrow}R^{\downarrow}\rangle)|\!\!\uparrow\downarrow\rangle,\\
  \frac{1}{\sqrt{2}}(|R^{\downarrow}R^{\downarrow}\rangle+|L^{\uparrow}L^{\uparrow}\rangle)|\!\!\downarrow\uparrow\rangle
  \rightarrow-\frac{1}{\sqrt{2}}(|L^{\uparrow}R^{\downarrow}\rangle+|R^{\downarrow}L^{\uparrow}\rangle)|\!\!\downarrow\uparrow\rangle.
  \end{split}
\end{equation}

\section{ENTANGLEMENT PURIFICATION FOR two-ELECTRON  systems}
In practice, the entanglement shared by two nodes in a quantum communication network is always less entangled, due to the channel noise involving in entanglement distribution process \cite{Quaninternet,nielsen2000}. The original maximal entangled state will probably become into a two-qubit mixed states due to the random character of the channel noise, both bit-flip and phase-flip errors take place randomly. Suppose two nodes, Alice and Bob, share an ensemble of less entangled two-qubit mixed states $\rho$. In order to share pairs of high-entangled electrons
they can perform purification to decrease both the bit-flip and the phase-flip errors.

\subsection{Purification for bit-flip error}

Suppose the ideal target electron pair connecting two nodes Alice and Bob are in state $|\phi^+_e\rangle=\frac{1}{\sqrt{2}}(|\uparrow\uparrow\rangle+|\downarrow\downarrow\rangle)$. When a bit-flip error takes place in entanglement distribution, the initial mixed entangled state ensemble shared
by  Alice and Bob will be  described by the density matrix,
\begin{equation}\label{cs}
  \begin{split}
  \rho=F|\phi^+_e\rangle\langle\phi_e^+|+(1-F)|\psi^+_e\rangle\langle\psi_e^+|.
  \end{split}
\end{equation}
Here $|\phi^+_e\rangle$ is the error-free component, and $|\psi^+_e\rangle$ is the component with bit-flip error $|\psi^+_e\rangle=\frac{1}{\sqrt{2}}(|\uparrow\downarrow\rangle+|\downarrow\uparrow\rangle)$. The coefficient $F$ is the probability obtaining the outcome $|\phi^+_e\rangle$  when measuring the electron pair in Bell basis $\{|\phi^{\pm}_e\rangle$,$|\psi^{\pm}_e\rangle\}$, where $|\phi^-_e\rangle=\frac{1}{\sqrt{2}}(|\uparrow\uparrow\rangle-|\downarrow\downarrow\rangle)$ and $|\psi^-_e\rangle=\frac{1}{\sqrt{2}}(|\uparrow\downarrow\rangle-|\downarrow\uparrow\rangle)$. It is identical to the fidelity of the mixed state $\rho$ with respect to target state  $|\phi^+_e\rangle$, which is defined as $F={tr(\rho|\phi^+\rangle\langle\phi^+|)}$.

\begin{figure}[!tpb]
  \centering
  \includegraphics[width=7.8cm]{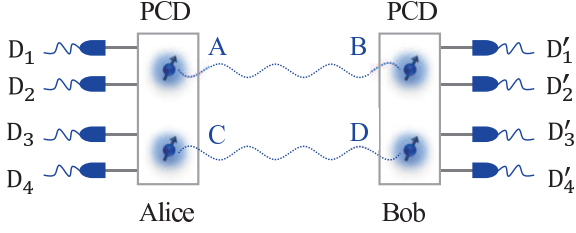}\\
  \caption{Schematic diagram for the principle of the EPP using PCD shown in Fig. \ref{qnd}. Here $D_i$ and $D_i'$ ($i=1,2,3,4$) are single-photon detectors. \emph{A} and \emph{C} represent electrons  in the node of Alice, \emph{B} and \emph{D} represent electrons   in the node of Bob.}\label{pro}
\end{figure}

The principle for bit-flip error purification is shown in Fig \ref{pro}. Suppose two entangled electron pairs $AB$ and $CD$ are in the same mixed state $\rho$. The combined state of the four electrons could be viewed as a mixture of four pure states $|\phi^{+}_e\rangle_{AB}|\phi^{+}_e\rangle_{CD}$, $|\phi^{+}_e\rangle_{AB}|\psi^{+}_e\rangle_{CD}$, $|\psi^{+}_e\rangle_{AB}|\phi^{+}_e\rangle_{CD}$, and $|\psi^{+}_e\rangle_{AB}|\psi^{+}_e\rangle_{CD}$, and the corresponding probabilities are $F^{2}$, $F(1-F)$, $F(1-F)$, and $(1-F)^{2}$, respectively.

To obtain an electron pair of higher fidelity, Alice and Bob  firstly perform a quantum nondemolition PCD respectively on the electron pair in their own nodes, that is they check the parity of their own electron-pair $AC$ and $BD$. If the four electron system is in state $|\phi^{+}_e\rangle_{AB}|\phi^{+}_e\rangle_{CD}$, and the corresponding probability $p_0=F^2$, the combined state of the photon-electron hybrid system before the two photons are measured will evolve into
\begin{eqnarray}\label{puri1}
\frac{1}{2}[(|\!\!\uparrow\uparrow\uparrow\uparrow\rangle+
  |\!\!\downarrow\downarrow\downarrow\downarrow\rangle)|\phi\rangle_a^p|\phi\rangle_b^p
+(|\!\!\downarrow\downarrow\uparrow\uparrow\rangle+
  |\!\!\uparrow\uparrow\downarrow\downarrow\rangle)|\psi\rangle_a^p|\psi\rangle_b^p].\nonumber\\
\end{eqnarray}
Here $|\phi\rangle^p_{a,b}=(|RR\rangle+|LL\rangle)/\sqrt{2}$
and $|\psi\rangle^p_{a,b}=(|RL\rangle+|LR\rangle)/\sqrt{2}$, and the subscripts \emph{a} and \emph{b} distinguish the photon pair belonging to Alice and Bob, respectively. If both of the two entangled photon pairs are in even parity $|RR\rangle$ ($|LL\rangle$), they will click the detectors of the same side in each of the two PCDs. Now, Alice and Bob will share a  four-electron entangled state
\begin{equation}\label{four}
  |\Phi_e^{+}\rangle=\frac{1}{\sqrt2}(|\!\!\uparrow\uparrow\uparrow\uparrow\rangle+
  |\!\!\downarrow\downarrow\downarrow\downarrow\rangle).
\end{equation}
However, if both of the two entangled photon pairs are in odd parity, $|RL\rangle$ ($|LR\rangle$), they will click the detectors in different sides of each PCD. The four-electron state shared by Alice and Bob will be
\begin{equation}\label{four2}
  |\Phi_o^{+}\rangle=\frac{1}{\sqrt2}(|\!\!\uparrow\uparrow\downarrow\downarrow\rangle+
  |\!\!\downarrow\downarrow\uparrow\uparrow\rangle).
\end{equation}

Subsequently, Alice and Bob perform a measurement on electron \emph{C} and \emph{D} in the basis $\{|\pm\rangle=(|\uparrow\rangle\pm|\downarrow\rangle)/{\sqrt2}\}$. If Alice and Bob both get the outcome $|+\rangle$ or $|-\rangle$, they will directly share a pair of maximally entangled electron pair \emph{AB} in state $|\phi^{+}_e\rangle$. However, if Alice and Bob get different outcomes, one gets $|+\rangle$ and the other gets $|-\rangle$, Alice and Bob need to perform a phase-flip operation on either electron \emph{A} or \emph{B} to share the target maximal entangled state $|\phi^{+}_e\rangle$.

For the component $|\psi^{+}_e\rangle_{AB}|\psi^{+}_e\rangle_{CD}$, the corresponding probability $p_1=(1-F)^2$, Alice and Bob will get the same outcome when performing parity-check measurement on their own electron pairs. However, when both of them get the even-parity outcome, the four electron state will be projected into
\begin{equation}\label{four3}
  |\Psi_e^{+}\rangle=\frac{1}{\sqrt2}(|\!\!\uparrow\downarrow\uparrow\downarrow\rangle+
  |\!\!\downarrow\uparrow\downarrow\uparrow\rangle).
\end{equation}
When both of them get the old-parity outcome, the four-electron state will be projected into
\begin{equation}\label{four4}
  |\Psi_o^{+}\rangle=\frac{1}{\sqrt2}(|\!\!\uparrow\downarrow\downarrow\uparrow\rangle+
  |\!\!\downarrow\uparrow\uparrow\downarrow\rangle).
\end{equation}
Now the measurement on  electrons \emph{C} and \emph{D} in the basis $\{|\pm\rangle\}$ will collapse both states $ |\Psi_e^{+}\rangle$ and $ |\Psi_o^{+}\rangle$ into the entangled state $|\psi^{+}_e\rangle_{AB}$, up to a local phase-flip feedback on either electron \emph{A} or \emph{B}. This leads to the error component that contributes to the final state after the first round purification and the corresponding probability is $p_1=(1-F)^2$.

For the other two components $|\psi^{+}_e\rangle_{AB}|\phi^{+}_e\rangle_{CD}$ and $|\phi^{+}_e\rangle_{AB}|\psi^{+}_e\rangle_{CD}$, Alice and Bob will never get the same parity outcome as they do in the first two cases ($|\phi^{+}_e\rangle_{AB}|\phi^{+}_e\rangle_{CD}$ and $|\psi^{+}_e\rangle_{AB}|\psi^{+}_e\rangle_{CD}$) when applying PCDs on their own electron pair \emph{AC} and \emph{BD}. The measurement on  electrons \emph{C} and \emph{D} projects the electron-pair \emph{AB} into a mixed state that consists of $|\phi^{+}_e\rangle_{AB}$ and $|\psi^{+}_e\rangle_{AB}$ with equal probability, which will be abandoned.

In practice, the two nodes, Alice and Bob, can distil a pair of higher fidelity electron-pair \emph{AB} in a nondeterministic way, by picking out the first two cases discussed above. The mixed state of electron-pair \emph{AB} now is described
by a density matrix similar to that shown in Eq. (\ref{cs}),
\begin{equation}\label{csn}
  \begin{split}
  \rho_1=F_1|\phi^+_e\rangle\langle\phi^+_e|+(1-F_1)|\psi^+_e\rangle\langle\psi^+_e|.
  \end{split}
\end{equation}
The fidelity of this new state $\rho_1$ is
\begin{equation}\label{f1}
F_1=\frac{F^{2}}{F^{2}+(1-F)^{2}}.
\end{equation}
It will be easily found that $F_1>F$ when the initial $F$ is larger than $1/2$, which means the probability of $|\phi^{+}_e\rangle$ in the mixed state $\rho_1$ is increased and the fidelity of $\rho_1$ is increased simultaneously. The efficiency of the EPP, which is defined as the probability that the original components are kept for partial measurement on electrons \emph{C} and \emph{D}, is
\begin{equation}\label{eta1}
\eta_1={F^{2}+(1-F)^{2}}.
\end{equation}
By iterating this purification process, Alice and Bob can in  principle correct the bit-flip error totally and share an ensemble of maximally entangled electron pairs in a heralded way.

\subsection{Purification for phase-flip error}
In the previous subsection, we detailed the purification process for bit-flip errors.
Now we start to explain the principle of the phase-flip error purification in the present scheme. Usually, the length variation of the transmission channel will inevitably introduce a random phase shift to the entangled electron pair. The dephasing noise will distort the initial maximal entangled state into a mixed one,
\begin{equation}\label{csn2}
  \begin{split}
  \rho_p=F|\phi^+_e\rangle\langle\phi^+_e|+(1-F)|\phi^-_e\rangle\langle\phi^-_e|,
  \end{split}
\end{equation}
where  $|\phi^-_e\rangle=\frac{1}{\sqrt{2}}(|\uparrow\uparrow\rangle-|\downarrow\downarrow\rangle$ represents the entangled state with a phase-flip error when compared with the target entangled state  $|\phi^+_e\rangle$.

Phase-flip errors, in practice, cannot be purified directly, while it can be converted into bit-flip errors. By performing the Hadamard rotations, $|\uparrow\rangle\rightarrow(|\uparrow\rangle+|\downarrow\rangle)/\sqrt{2}$ and $|\downarrow\rangle\rightarrow(|\uparrow\rangle-|\downarrow\rangle)/\sqrt{2}$  on each electron, $|\phi^+_e\rangle$ will be kept unchanged under these rotations, while $|\psi^-_e\rangle$ evolves into the state $|\psi^+_e\rangle=\frac{1}{\sqrt{2}}(|\uparrow\downarrow\rangle+|\downarrow\uparrow\rangle$ that contains a bit-flip error. In this way, Alice and Bob turn the phase-flip error  into a bit-flip error, and then they can purify the bit-flip error with the protocol presented in previous subsection.

 \begin{figure}[tpb]
  \centering
  \includegraphics[width=8.5cm]{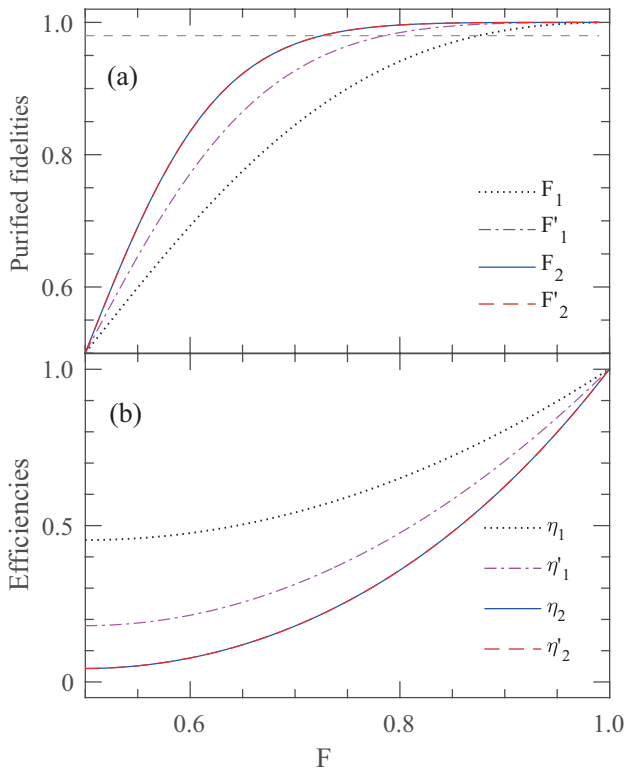}
  \caption{(a) The purified fidelities of different purification arrangements versus the initial fidelity $F$. Here $F_1$ denotes the purified fidelity after the first round purification; $F'_1$ denotes the purified one using two electron pairs of fidelities $F_1$ and $F$; $F_2$ denotes the purified one using two identical electron pairs of fidelities $F_1$; $F'_2$ denotes the one using two electron pairs of fidelities $F'_1$ and $F$; The short-dashed line represents a threshold fidelity of $F_{th}=0.98$. (b) The efficiencies of different purification arrangements versus the initial fidelity $F$. Each process is denoted with the same lines as that in (a).}\label{fig4}
\end{figure}

\subsection{Purification with non-identical electron pairs}

In quantum networks, quantum error correction with a given redundant encoding can work efficiently when the fidelity of electron pairs are higher than a  critical threshold value \cite{Qthresh1,Qthresh2}. Although one can iterate the purification protocol with two copies of identical electron pairs, and then generate an entangled electron pair of a higher fidelity, it might be more economic to purify electron pairs by assisting electron pairs with less fidelity from the point of practice. One can purify an electron pair generating from the \emph{m}-th purification round with an electron pair from {(\emph{m-n})}-th purification round with $n\leq m$.

After the first round of purification for the bit-flip errors, electron pairs will evolve into a mixed state with the fidelity $F_1=\frac{F^{2}}{F^{2}+(1-F)^{2}}$. Instead of performing purification with two identical electron pairs with fidelity $F_1$ \cite{EPPpan01,EPPpan2,EPPSheng1,EPPpdeng1}, Alice and Bob perform the same purification procedure on two electron pairs ${A'B'}$ and ${C'D'}$ with fidelities $F_1$ and $F$, respectively.
By selecting  the identical parity on electron pairs ${A'C'}$ and ${B'D'}$ and then performing the measurement on electrons ${C'}$ and ${D'}$, the electron pair ${A'B'}$ will be projected into a new state $\rho'_1$,
\begin{equation}\label{rho3}
  \begin{split}
  \rho'_1=F_1'|\phi^+_e\rangle\langle\phi^+|+(1-F_1')|\phi^-_e\rangle\langle\phi^-|,
  \end{split}
\end{equation}
up to a local phase-flip operation on either electron ${A'}$ or ${B'}$. The fidelity $F'_1$ of the new mixed state $\rho'$ is
\begin{eqnarray}
  F_1' &=& \frac{F^3}{F^3+(1-F)^3}.
\end{eqnarray}
The corresponding efficiency of the EPP process is
\begin{eqnarray}
  \eta_1' &=& F^3+(1-F)^3,
\end{eqnarray}
which equals the success probability of two purification procedures: one is the creation of electron pair ${A'B'}$ with the purification procedure using two electron pairs of initial fidelity $F$, and the other one is the purification procedure using  electron pairs ${A'B'}$ and  ${C'D'}$ with initial fidelity $F'$ and $F$.

In a similar calculation, one can give out the fidelity
$F_2$ of the second round purification using two pair of electrons
with initial fidelity $F_1$.
\begin{eqnarray}
  F_2 = \frac{F^4}{F^4+(1-F)^4}.
\end{eqnarray}
The corresponding efficiency $\eta_2$ is the joint probability of
the success of two first round purifications with initial fidelity
$F$ and one second round purification with initial fidelity $F_1$,
\begin{eqnarray}
   \eta_2= F^4+(1-F)^4.
\end{eqnarray}
In addition, one can also find the fidelity $F_2'=F_2$ and the efficiency $\eta_2'=\eta_2$ of the purification protocol using two electron pairs of different initial fidelity, say $F'_1$ and $F$.  These two different purification arrangements consume the same quantum source and the corresponding performances are shown in Fig. \ref{fig4}.

 \begin{figure*}[!tpb]
  \centering
  \includegraphics[width=15cm ]{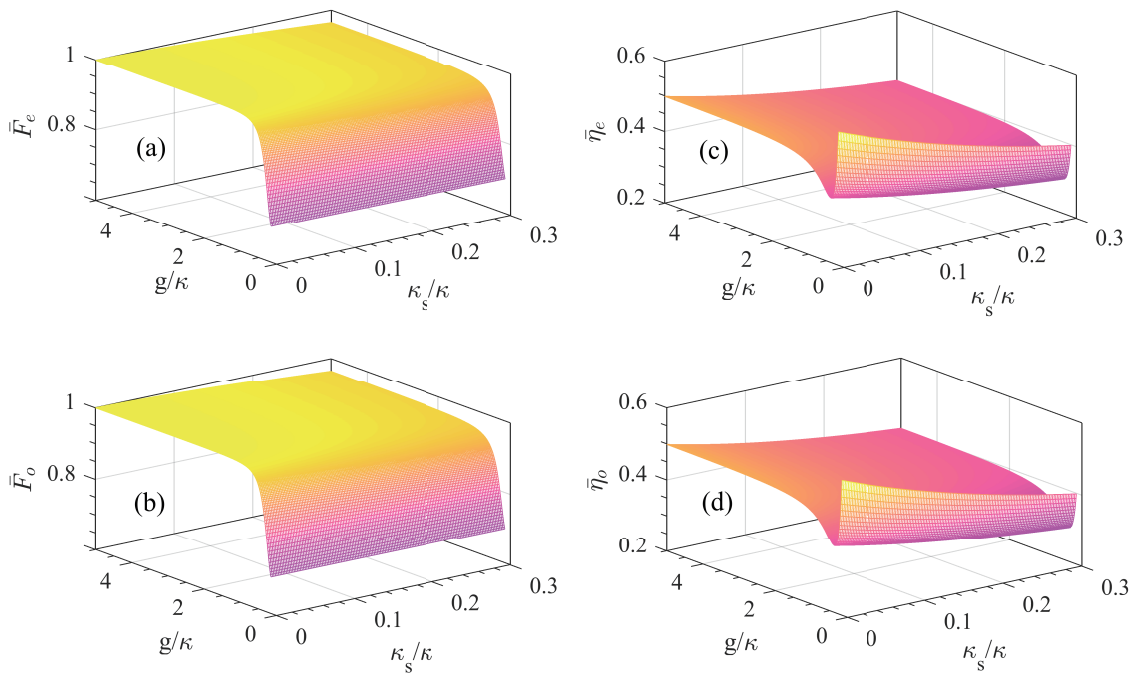}
  \caption{(a) The average fidelity and efficiency of our PCD scheme vs the side-leakage rate $\kappa_s/\kappa$ and the QD-cavity coupling rate $g/\kappa$ with $\gamma/\kappa=0.1$. (a) $\bar{F}_e$ is the average fidelity for the even outcome of the PCD; (b) $\bar{F}_o$ is the average fidelity for the odd outcome; (c) $\bar{\eta}_e$ is the average efficiency for the even outcome; (d) $\bar{\eta}_o$ is the average efficiency for the odd outcome.}\label{fig5}
\end{figure*}

\subsection{Performance of entanglement purification}
The purified fidelities and efficiencies versus the initial fidelity are shown in Fig. \ref{fig4}. For the purified fidelity of the final entangled electron pair, it is only dependent on the number of the electron pair that is used for purification, while it is independent on the purification arrangements. When four electron pairs are used to perform purification to generate a high-fidelity electron pair.  The purified fidelity $F_2$ of second round purification is identical to $F_2'$ which is obtained by purifying an electron pair of a fidelity $F$ three times with three identical electron pairs of the fidelity $F$, shown in \ref{fig4} (a). Moreover, in order to share entangled channel of a fidelity higher than a threshold value, such as $F_{th}=0.98$, Alice and Bob could choose different purification arrangements according to the initial fidelity of electron-pair ensembles available. For ensembles with the initial fidelity $F>0.724$ and $F<0.785$, it will be more efficient to perform the symmetrical purification to the second round to generate the desired electron pairs. However, for ensembles with the initial fidelity $F>0.785$ and $F<0.875$, it will be enough to generate electron pairs of a fidelity $F'_1>F_{th}$ by performing purification using electron pairs with fidelities $F_1$ and $F$. It will consume less electron pairs when compared to the second round purification in symmetrical arrangement. For ensembles with initial fidelity $F>0.875$ and $F<0.98$, one just needs to perform the purification once to get the desired electron pairs.

The efficiencies of different purification arrangement are shown in Fig. \ref{fig4} (b).
The efficiencies are independent on the purification order as well,
and they are significantly affected by the initial fidelity $F$ and
the number of electron pairs consuming in purification processes.
Therefore, it will be more economic to use the symmetrical purification arrangement for low-entangled ensembles at first, and then
perform purification with electron pairs generating from different purification rounds, such as the one $F>0.785$ and $F<0.875$, to generate the final desired electron pairs. In this way, one can perform the purification with a higher efficiency and simultaneously consume less quantum sources.

\section{Practical PCD with current experimental parameters}
The single photon scattering involving in our PCD, in practice, will be more complicated than the simplified version shown in Eq. (\ref{mcf}) for addressing the PCD explicitly. A single photon will be scattered into both the reflection and transmission modes simultaneously with different probability amplitudes. Therefore, for a given input state, the practical output state after the scattering process should be modified into \cite{QR1,QR2}
\begin{eqnarray}
\begin{split}
|R^{\uparrow},\uparrow\rangle \;\; \rightarrow  \;\;&
r|L^{\downarrow},\uparrow\rangle+t|R^{\uparrow},\uparrow\rangle \\
|R^{\downarrow},\uparrow\rangle  \;\; \rightarrow  \;\;&  t_0|R^{\downarrow},\uparrow\rangle+r_0|L^{\uparrow},\uparrow\rangle, \\
|L^{\downarrow},\uparrow\rangle   \;\; \rightarrow  \;\;&
r|R^{\uparrow},\uparrow\rangle+t|L^{\downarrow},\uparrow\rangle, \\
|L^{\uparrow},\uparrow\rangle  \;\; \rightarrow  \;\;&
t_0|L^{\uparrow},\uparrow\rangle+r_0|R^{\downarrow},\uparrow\rangle,
\label{transuppp}
\end{split}
\end{eqnarray}
and
\begin{eqnarray}
\begin{split}
|R^{\uparrow},\downarrow\rangle  \;\; \rightarrow  \;\;&
t_0|R^{\uparrow},\downarrow\rangle+r_0|L^{\downarrow},\downarrow\rangle, \\
|R^{\downarrow},\downarrow\rangle  \;\; \rightarrow  \;\;&
r|L^{\uparrow},\downarrow\rangle+t|R^{\downarrow},\downarrow\rangle, \\
|L^{\uparrow},\downarrow\rangle  \;\; \rightarrow  \;\;&
r|R^{\downarrow},\downarrow\rangle+t|L^{\uparrow},\downarrow\rangle, \\
|L^{\downarrow},\downarrow\rangle  \;\; \rightarrow  \;\;&
t_0|L^{\downarrow},\downarrow\rangle+r_0|R^{\uparrow},\downarrow\rangle.
\label{transdownpp}
\end{split}
\end{eqnarray}
Here the coefficients $r$ ($t$) and $r_0$ ($t_0$) are the reflection (transmission)
coefficients for hot and cold cavities \cite{QR1,QR2,QD2,QD.2}, respectively.
\begin{eqnarray}   
\begin{split}
r\;\;=\;\;&\frac{\frac{\kappa_s}{2}+\frac{2g^{2}}{\gamma}}{\kappa+\frac{\kappa_s}{2}+\frac{2g^{2}}{\gamma}}, \\
t\;\;=\;\;&\frac{-\kappa}
{\kappa+\frac{\kappa_s}{2}+\frac{2g^{2}}{\gamma}},
 \label{hotrt}
\end{split}
\end{eqnarray}
and
\begin{eqnarray}   
\begin{split}
r_0\;\;=\;\;&\frac{\frac{\kappa_s}{2}}{\kappa+\frac{\kappa_s}{2}}, \\
t_0\;\;=\;\;&\frac{-\kappa}
{\kappa+\frac{\kappa_s}{2}},
 \label{coldrt}
\end{split}
\end{eqnarray}
where the input photon is resonant to the QD transition and the cavity mode. $\kappa$, $\kappa_s$, and $\gamma/2$  are the cavity-field decay
rate, the cavity side-leakage rate, and   the QD decay rate, respectively. $g$ is the coupling
strength between the QD and the cavity mode.

The fidelity of a PCD is defined as the overlap between the outputs for the ideal and the practical scattering, and the corresponding efficiency is defined as the probability that the PCD succeeds, which is heralded by two-photon detection in our PCD scheme. Here we exploited the average fidelity and efficiency over all pure states to test the performance of our PCD \cite{Nielsen1}, shown in Fig. \ref{fig5}. The average fidelity $\bar{F}_e$ ($\bar{F}_o$) for the even (odd) outcome of the PCD is larger than $0.990$ for $\kappa_s/\kappa<0.1$  and $g/\kappa>1.5$. Meanwhile, the corresponding  average efficiency $\bar{\eta}_e$ ($\bar{\eta}_o$) is larger than $0.439$. All the average fidelities and efficiencies increase when the QD-cavity coupling $g/\kappa$ is increased or the cavity side-leakage $\kappa_s/\kappa$ is decreased. Currently, the strong coupling $g/(\kappa+\kappa_s)>1$ has been observed in various QD-cavity systems \cite{QDStron1,QDStron2,QDStron3}. Furthermore, one can control $g$ and $\kappa$ independently to achieve a larger ratio $g/(\kappa+\kappa_s)$ \cite{tuneRatio}, since the coupling strength $g$ depends on the dipole of QD exciton and the cavity volume, while the cavity-field
decay rate $\kappa$ is determined by the cavity quality. Therefore, our PCD scheme   works faithfully with a near-unity fidelity, and the total average efficiency will be larger than $0.878$, which could, in principle, approach unity when ideal single-photon scattering is used.

\section{Discussion and summary}


Usually, channel noise will inevitably decrease the fidelity of the entanglement, which  decreases the security of quantum communication and increases the probability of error output in distributed quantum networks \cite{QNsecure1,QNsecure2}. Entanglement purification is an efficient passive method to depress the negative effect of channel noise and to improve the fidelity of less entangled systems \cite{EPP1B}. Here, we described an efficient protocol for entanglement purification for electron-spin systems by constructing a nondestructive PCD with an entangled photon pair and rearranging the order of purification process. In our PCD, each photon will only be scattered by one QD-cavity union and then be detected directly after it leaves the cavity. Therefore, our PCD works in a parallel style other than a cascaded one \cite{QD22,QDpcd1}. Moreover, the photon is encoded in polarization degree of freedom and the phase fluctuation due to the variation of optical-path length will appear as a global  coefficient \cite{polarization}. In addition, no single-photon quantum interference is involved in our PCD scheme, since the photons are detected directly after scattered by the QD-cavity unions. Therefore, our PCD will, to some extend, be robust to the fluctuation of optical-path length \cite{Quaninternet} between the entangled photon source and the QD-cavity union and it could be directly extended to multiple-particle systems \cite{Mult00,Mult0,Mult1,Mult2}.

By rearranging the order of purification process, we find that both the fidelity and the efficiency are determined by the initial fidelity and the number of the electron pairs, while they are independent on the purification orders. Therefore, it will be more economic to perform purification in a combination style with both symmetrical purification using identical electron pairs and purification using electron pairs from different purification rounds. In this way, we can purify a given electron pair with less ancillary electron pairs and simultaneously a higher efficiency when compared with the  symmetrical purification protocol using only identical electron pairs \cite{EPP1B,Qerrer,EPPpan01,EPPpan2,EPPSheng1,EPPpdeng1,QDpcd1}.

In summary, we have proposed an efficient PCD for two electrons based on double-sided microcavity and polarization-entangled photon  pair. The PCD works in a parallel style and is robust to the phase fluctuation of the optical path length, which is more efficient for nonlocal quantum information process and then large scale quantum computing. Based on this, we gave out an economic entanglement purification protocol. We found that the fidelity of the target electron pair is only determined by both the number and the initial fidelity of ancillary electron pairs, while it is independent on the purification orders. Our protocol will find its applications in the future quantum communication and distributed quantum networks.

\section*{ACKNOWLEDGMENTS}

This work was supported by the National Natural Science Foundation
of China under Grants No. 11674033, No.  11474026, and No. 11505007,
and the Fundamental Research Funds for the Central Universities
under Grant No. 2015KJJCA01.


\begin{thebibliography}{199}
\bibitem{Quaninternet}
H. J. Kimble,   Nature, \textbf{453}, 1023-1030 (2008).

\bibitem{nielsen2000}
M. A. Nielsen and  I. L. Chuang, \emph{Quantum computation and quantum
information} (Cambridge University Press, Cambridge, 2010).

\bibitem{QMee}
V. Giovannetti, S. Lloyd, and L. Maccone,   Nature Photon. \textbf{5}, 222 (2011). 

\bibitem{DisQC1}
L. Jiang, J. M. Taylor, A. S. S{\o}rensen, and  M. D. Lukin,  Phys. Rev.  A \textbf{76}, 062323 (2007).

\bibitem{DisQC2}
 W. D\"{u}r, and H. J. Briegel,  Phys. Rev. Lett. \textbf{90}, 067901 (2003).



\bibitem{DisQC3}
W. Qin, X. Wang, A. Miranowicz, Z. Zhong, and F.  Nori, Phys. Rev. A \textbf{96}, 012315 (2017).


\bibitem{DisQC33}
Y. H. Kang,  Z. C. Shi, B. H. Huang,  J.  Song, and  Y. Xia,  Ann.
Phys. (Berlin) \textbf{529}, 1700154  (2017).



\bibitem{DisQC4}
Y. B. Sheng and L.  Zhou,    Sci. Bull. \textbf{62}, 1025 (2017).


\bibitem{ekert1991}
A. K. Ekert, Phys. Rev. Lett. \textbf{67}, 661 (1991).

\bibitem{bennett1992a}
C. H. Bennett  and  S. J. Wiesner, Phys. Rev. Lett. \textbf{69}, 2881 (1992).


\bibitem{LiaoPanQKD}
X. H. Li,  F. G. Deng,  and H. Y. Zhou, Phys. Rev. A \textbf{78},
022321 (2008).


\bibitem{Lo2014QKD}
 H. K. Lo, M. Curty, and K. Tamaki,  Nature Photon. \textbf{8}, 595 (2014).






\bibitem{telep1}
C. H. Bennett,  G. Brassard, C. Crepeau, R. Jozsa, A. Peres, and  W. K. Wootters,    Phys.  Rev.
Lett. \textbf{70}, 1895 (1993).




\bibitem{RenPanteleportation}
J. G. Ren, P. Xu, H. L. Yong, L. Zhang, S. K. Liao, and J. Yin,  \emph{et al},    Nature, \textbf{549}, 70 (2017).


\bibitem{QSS0}
 R. Cleve, D.  Gottesman, and H. K.  Lo,    Phys. Rev. Lett. \textbf{83}, 648 (1999).

\bibitem{QSS1}
M. Hillery, V. Bu\u{z}ek, and  A. Berthiaume,  Phys. Rev. A \textbf{59}, 1829 (1999).

\bibitem{QSS2}
 A. M. Lance, T. Symul, W. P.  Bowen, B. C. Sanders, and P. K. Lam,   Phys.
Rev. Lett. \textbf{92}, 177903 (2004).




\bibitem{QSDC1}
G. L. Long and X. S. Liu,   {Phys. Rev. A} \textbf{65},
032302 (2002).

\bibitem{QSDC2}
F. G. Deng, G. L.  Long,  and X. S. Liu,
{Phys. Rev. A}  \textbf{68}, 042317 (2003).

\bibitem{QSDC3}
F. G. Deng and G. L. Long,   Phys. Rev. A \textbf{69}, 052319 (2004).


\bibitem{QSDC6}
J. Y. Hu, B. Yu,  M. Y. Jing,  L. T.  Xiao, S. T. Jia,  G. Q. Qin, and G. L. Long,   Light Sci. Appl. \textbf{5}, e16144 (2016).

\bibitem{QSDC5}
W. Zhang,  D. S. Ding,  Y. B. Sheng, L. Zhou, B. S. Shi, and G. C. Guo,   Phys. Rev. Lett. \textbf{118}, 220501 (2017).

\bibitem{QSDC7}
F. Zhu, W. Zhang,  Y. B. Sheng, and  Y. Huang,   Sci. Bull. \textbf{62}, 1519 (2017).


\bibitem{QR0}
H. J. Briegel, W. D\"{u}r, J. I. Cirac, and P.  Zoller,   Phys. Rev.  Lett. \textbf{81}, 5932 (1998).

\bibitem{QR1}
T. J. Wang,  S. Y. Song, and G. L.  Long,  Phys. Rev. A 85, 062311 (2012).

\bibitem{QR2}
T. Li, G. J. Yang, and F. G. Deng,   Phys. Rev.  A \textbf{93}, 012302 (2016).



\bibitem{QS0}
 M. Zukowski, A. Zeilinger, M. A. Horne, and A. K. Ekert,   Phys. Rev. Lett. \textbf{71}, 4287 (1993).

\bibitem{QS1}
X. Su, C. Tian,  X. Deng,  Q. Li, C. Xie, and K. Peng,  Phys. Rev. Lett. \textbf{117}, 240503 (2016).

\bibitem{QS2}
L. Zhou and Y. B. Sheng, Phys. Rev. A \textbf{92}, 042314 (2015).

\bibitem{QS3}
C. M. Xie, Y. M. Liu, J. L. Chen, X. F. Yin, and Z. J. Zhang,  Sci. China-Phys. Mech. Astron. \textbf{59}, 100314 (2016).


\bibitem{ECP1B}
C. H. Bennett, H. J. Bernstein,  S. Popescu, and  B. Schumacher,   Phys. Rev.  A, \textbf{53}, 2046 (1996).

\bibitem{ECP1add1} Y. B. Sheng, F. G. Deng, and H. Y. Zhou,  Phys. Rev. A \textbf{77}, 062325 (2008).


\bibitem{ECP1add2}
B.C. Ren, F. F. Du, and F. G. Deng,   Phys. Rev. A \textbf{88},
012302 (2013).



\bibitem{Eccs1}
Y. B.  Sheng, L. Zhou, S. M. Zhao, and B. Y. Zheng, Phys. Rev. A \textbf{85}, 012307   (2012).


\bibitem{Eccs2}
Y. B.  Sheng, L. Zhou, and S. M. Zhao, Phys. Rev. A  \textbf{85}, 042302 (2012).


\bibitem{ECP1add3} F. G. Deng,   Phys. Rev. A \textbf{85},
022311 (2012).


\bibitem{Eccl1}
X. H. Li and S. Ghose,  Phys. Rev. A \textbf{91}, 062302 (2015).


\bibitem{Eccl2}
X. H. Li and S. Ghose,  Opt. Express \textbf{23}, 3550 (2015).



\bibitem{Qerrer}
W. D\"{u}r and H. J. Briegel,  Rep. Prog. Phys. \textbf{70}, 1381 (2007).


\bibitem{EPP1B}
C. H. Bennett, G. Brassard, S. Popescu, B. Schumacher, J. A. Smolin, and W. K.  Wootters,   Phys. Rev.  Lett. \textbf{76}, 722 (1996).



\bibitem{EPPpan01}
J.W. Pan, C. Simon, and  A.  Zellinger,   Nature,
\textbf{410}, 1067 (2001).

\bibitem{EPPpan2}
C. Simon and J.W.  Pan,    Phys. Rev.
Lett. \textbf{89}, 257901 (2002).

\bibitem{EPPSheng1}
Y. B. Sheng,  F. G. Deng, and H. Y. Zhou,    Phys. Rev. A \textbf{77}, 042308 (2008).

\bibitem{EPPpdeng1}
 F. G. Deng,   Phys. Rev. A \textbf{84}, 052312 (2011).

\bibitem{DEPP1}
Y. B. Sheng and F. G Deng,   Phys. Rev. A \textbf{81}, 032307 (2010).



\bibitem{DEPP2}
X. H. Li,   Phys. Rev.
A \textbf{82}, 044304 (2010).

\bibitem{DEPP3}
Y. B. Sheng and F. G Deng,   Phys. Rev. A \textbf{82}, 044305
(2010).

\bibitem{DEPP4}
F. G.  Deng,     Phys.  Rev.
A \textbf{83}, 062316 (2011).


\bibitem{DEPP5}
 Y. B. Sheng and L. Zhou,   Laser Phys. Lett. \textbf{11}, 085203  (2014).

\bibitem{EPPsl2}
Y. B. Sheng and L. Zhou, Sci. Rep. \textbf{5}, 7815 (2015).


\bibitem{HEPP1}
 B. C. Ren and F. G.  Deng,  Laser Phys. Lett. \textbf{10}, 115201 (2013).

\bibitem{HEPP2}
 B. C. Ren, F. F. Du, and F. G.  Deng,   Phys. Rev. A \textbf{90}, 052309 (2014).

\bibitem{HEPP3}
 G. Y. Wang, Q. Liu, and F. G. Deng,   Phys. Rev. A \textbf{94}, 032319 (2016).



\bibitem{EPPHybrid}
Y. B. Sheng, L. Zhou, and G. L. Long,   Phys. Rev. A \textbf{88}, 022302 (2013).


\bibitem{EPPKerr}
C. Wang, Y. Zhang, and G. S. Jin,   Quantum Inf. Comput. \textbf{11}, 0988 (2011).


\bibitem{DEPPw1}
 C. Cao, C. Wang,  L. Y. He, and R. Zhang,    Opt. Express, \textbf{21}, 4093 (2013).

\bibitem{DEPPw2}
L. Zhou and Y. B. Sheng, Sci. Rep. \textbf{6}, 28813 (2016).

\bibitem{DEPPw3}
 T. J. Wang,  L. L. Liu,  R. Zhang, C. Cao, and  C. Wang,   Opt. Express. \textbf{23}, 9284 (2015).

\bibitem{DEPPw4}
T. J. Wang,  S. C. Mi, and C.  Wang,    Opt. Express. \textbf{25}, 2969  (2017).


\bibitem{QD00}
D. Loss and  D. P. DiVincenzo,    Phys. Rev.  A \textbf{57}, 120 (1998).

\bibitem{QD001}
I. Buluta, S. Ashhab, and F. Nori,   Rep. Prog. Phys. \textbf{74}, 104401  (2011).


\bibitem{QD02}
R. B. Liu,  W.  Yao, and L. J. Sham,    Adv. Phys. \textbf{59}, 703 (2010).

\bibitem{QD01}
R. J. Warburton,    Nat. Mater. \textbf{12}, 483 (2013).

\bibitem{QD03}
P. Lodahl, S. Mahmoodian, and S. Stobbe,    Rev. Mod. Phys. \textbf{87}, 347  (2015).

\bibitem{QD04}
D. Press,  K. De Greve, P. L. McMahon, T. D. Ladd, B. Friess,  C. Schneider, M. Kamp,  S. H\"{o}ing,   A. Forchel, and  Y. Yamamoto,   Nature Photon. \textbf{4}, 367 (2010).


\bibitem{QD1}
C. Y. Hu, A. Young, J. L. O'Brien,  W. J. Munro, and  J. Rarity,    Phys. Rev.  B, \textbf{78}, 085307 (2008).

\bibitem{QD2}
C. Y. Hu, W. J. Munro, J. L. O'Brien, and J. G. Rarity,  Phys. Rev.  B \textbf{80}, 205326 (2009).

\bibitem{QD.2}
 C. Bonato, F. Haupt,  S. S. Oemrawsingh,  J. Gudat,  D. Ding,  M. P. van Exter, and  D. Bouwmeester,    Phys. Rev. Lett. \textbf{104}, 160503 (2010).

\bibitem{QD22}
 A. B. Young,  C. Y. Hu, and J. G. Rarity,     Phys. Rev.  A \textbf{87}, 012332 (2013).

\bibitem{QD3}
H. R. Wei and F. G.  Deng,    Opt. Express. \textbf{21}, 17671 (2013).

\bibitem{QD42}
B. C. Ren, H. R. Wei, and F. G. Deng, Laser Phys. Lett. \textbf{10},
095202 (2013).

\bibitem{QD41}
B. C. Ren and  F. G. Deng, Sci. Rep. \textbf{4},  4623 (2014).


\bibitem{QD4}
H. F. Wang,  A. D. Zhu,  S. Zhang, and K. H. Yeon,    Phys. Rev.  A
\textbf{87}, 062337 (2013).


\bibitem{QDreview} F. G. Deng, B. C. Ren, and X. H. Li, Sci. Bull. \textbf{62}, 46
(2017).



\bibitem{QD50}
C. Y. Hu and J. G.  Rarity,    Phys. Rev.  B \textbf{ 83}, 115303 (2011).

\bibitem{QD5}
 B. C. Ren, H. R. Wei, M. Hua, T. Li, and F. G. Deng,  Opt. Express \textbf{20}, 24664
 (2012).


\bibitem{QD5add1}
 T. J. Wang, Y. Lu, and G. L. Long,  Phys. Rev. A \textbf{86}, 042337
 (2012).



\bibitem{QD5add2}
G. Y. Wang, Q. Ai, B.C. Ren, T. Li, and F.G. Deng,  Opt. Express
\textbf{24}, 28444 (2016).







 \bibitem{QDrout1}
C. Y.  Hu, Sci. Rep. \textbf{7}, {45582}  (2017).

\bibitem{QDrout2}
C.  Cao,  Y. W.  Duan,  X.  Chen,  R.  Zhang,  T. J.  Wang, and  C.  Wang,  Opt. Express \textbf{25},  {16931} (2017).


\bibitem{QDpcd1}
 C. Wang, Y. Zhang, and  G. S.   Jin,    Phys. Rev.  A \textbf{84}, 032307 (2011).

\bibitem{QDpcd2}
 T. Li, and  F. G. Deng,  Phys. Rev.  A \textbf{94}, 062310  (2016).

\bibitem{Qthresh1}
J. I. Cirac, A. K. Ekert, S. F. Huelga, and C.  Macchiavello,   Phys. Rev. A \textbf{59}, 4249 (1999).

\bibitem{Qthresh2}
 Y. Li and S. C. Benjamin,    New J. Phys. \textbf{14}, 093008 (2012).

\bibitem{Nielsen1}
M. A. Nielsen,    Phys. Lett.
A \textbf{303}, 249  (2002).

\bibitem{QDStron1}
J. P. Reithmaier, G. S\c{e}k, A. L\"{o}ffler, C. Hofmann, S. Kuhn, S. Reitzenstein, L. V. Keldysh, V. D. Kulakovskii, T. L. Reinecke, and A. Forchel,
  Nature, \textbf{432}, 197 (2004).


\bibitem{QDStron2}
S. Kreinberg, W. W. Chow, J. Wolters,  C.  Schneider,  C. Gies,  F. Jahnke,  S. H\"{o}fling,  M.  Kamp, and  S. Reitzenstein,   Light Sci. Appl. \textbf{6}, e17030 (2017).

\bibitem{QDStron3}
P. Senellart, G.  Solomon, and A. White,  Nat. Nanotechnol. \textbf{12}, 1026  (2017).

\bibitem{tuneRatio}
 S. Reitzenstein and A. Forchel,   J. Phys. D. Appl. Phys. \textbf{43}, 033001 (2010).

\bibitem{QNsecure1}
 V. Scarani,  H. Bechmann-Pasquinucci,  N. J. Cerf, M. Du\u{s}ek,  N. L\"{u}tkenhaus, and  M. Peev,  Rev. Mod. Phys. \textbf{81}, 1301 (2009).

\bibitem{QNsecure2}
N. H. Nickerson,   J. F.  Fitzsimons,  and S. C. Benjamin,    Phys.  Rev.  X  \textbf{4},  041041  (2014).

\bibitem{polarization}
B. Zhao, Z. B. Chen, Y. A. Chen, J. Schmiedmayer, and J. W, Pan, Phys. Rev. Lett. \textbf{98}, 240502 (2007).

\bibitem{Mult00}
P. B. Li and F. L. Li, Opt. Express \textbf{19}, 1207 (2011).

\bibitem{Mult0}
S. Ashhab, P.  C.  De Groot, and F.  Nori, Phys. Rev. A \textbf{85}, 052327 (2012).

\bibitem{Mult1}
B.C. Ren, G.Y. Wang, F.G. Deng,  Phys. Rev.A \textbf{91},032328
(2015).




\bibitem{Mult2}
L. Zhou and Y. B. Sheng, Ann. Phys. \textbf{385}, 10 (2017).

\end{thebibliography}
\end{document}